\documentclass[twocolumn,aps,prb,showpacs,superscriptaddress]{revtex4}

\usepackage{amsmath}
\usepackage{amssymb}
\usepackage{epsfig}
\usepackage{color}

\newcommand{\be}{\begin{equation}}
\newcommand{\ee}{\end{equation}}
\newcommand{\nn}{\nonumber \\}
\newcommand{\ij}{\langle ij \rangle}

\newcommand{\ba}{\begin{eqnarray}}
\newcommand{\ea}{\end{eqnarray}}

\newcommand{\abar}{a^{\dag}}
\newcommand{\cbar}{c^\dag}
\newcommand{\dbar}{d^\dag}

\newcommand{\betabar}{\beta^\dag}
\newcommand{\Bbar}{B^\dag}
\newcommand{\Cbar}{C^\dag}
\newcommand{\Ubar}{U^*}
\newcommand{\gammabar}{\gamma^\dag}

\newcommand{\psibar}{\psi^{\dag}}

\newcommand{\bw}{\begin{widetext}}
\newcommand{\ew}{\end{widetext}}

\newcommand{\bpm}{\begin{pmatrix}}
\newcommand{\epm}{\end{pmatrix}}

\newcommand{\bbar}{b^\dag}

\newcommand{\Deltabar}{\Delta^*}

\begin{document}
\title{Exciton formation in graphene bilayer}

\author{Raoul Dillenschneider}
\email[Electronic address:$~~$]{raoul.dillenschneider@physik.uni-augsburg.de}
\affiliation{Department of Physics, University of Augsburg, Germany}

\author{Jung Hoon Han}
\email[Electronic address:$~~$]{hanjh@skku.edu}
\affiliation{Department of Physics, BK21 Physics Research Division,
Sungkyunkwan University, Suwon 440-746, Korea} \affiliation{CSCMR,
Seoul National University, Seoul 151-747, Korea}
\date{\today}

\begin{abstract}
Exciton instability in graphene bilayer systems is studied in the
case of a short-ranged Coulomb interaction and a finite voltage
difference between the layers. Self-consistent exciton gap
equations are derived and solved numerically and analytically under
controlled approximation. We obtain that a critical strength of the
Coulomb interaction exists for the formation of excitons.
The critical strength depends on the amount of voltage difference
between the layers and on the inter-layer hopping parameter.
\end{abstract}

\pacs{71.35.-y,71.20.Mq,78.67.Pt}

\maketitle

\section{Introduction}

Graphene, layers of two-dimensional honeycomb-array of carbon atoms, has
attracted much interest these last few years due to its recent experimental
accessibility \cite{graphene-QH_P1,graphene-QH_P2,graphene-QH_P3} and a wide
variety of interesting properties
\cite{Geim_Novoselov_2007,McCannFalko, AbergelFalko}. Both the
single-layer and the multi-layer graphenes are studied intensely.
Much of the peculiar properties of the graphene layers arises from
the energy spectrum near the so-called Dirac nodal points and
the non-trivial topological structure of the wave functions around
them \cite{Semenoff,Novoselov_Nature438}.

As the engineering application of the graphene layers attracts
increasing significance, we need to explore, experimentally and
theoretically, ways to enrich graphene's electrical properties and
to control them. One way to achieve some control over the electrical
properties is to change the number of layers and/or the bias applied
across the layers. A recent experimental realization of the biased
graphene bilayer is such an example\cite{Castro,Castro2,OOstinga}. By
applying a gate bias across the two-layered graphene, the authors of
Refs. \onlinecite{Castro} and \onlinecite{OOstinga} have observed a
tunable energy gap varying with the bias (see Fig.
\ref{fig3:Spectrum} for the bilayer graphene energy bands in the
presence of bias). The bias can also potentially control the
formation of excitons. Since the applied bias leads to the charge
imbalance in the two layers, it is natural to suspect that the
Coulomb attraction of the excess electrons and holes on opposite
layers would lead to an exciton instability similar to the
situations considered in an earlier
literature\cite{balatsky,exciton-conductivity,tJ-exciton}. If so, it
will provide an additional control over the graphene as the
formation of excitons is known to affect the electrical properties
significantly \cite{halperin-rice-review,exciton-conductivity}.

Recent works on the exciton instability in a single-layer graphene
are based on the Dirac Hamiltonian description
\cite{Khveshchenko_2001,Khveshchenko_Shively}. The exciton gap is
derived and solved through a self-consistent equation similar to the
one appearing in the chiral symmetry breaking phenomenon
\cite{Appelquist}. It was shown that an exciton can be formed under
a strong long-ranged particle-hole
interaction\cite{Leal_Khveshchenko}. Exciton can also be formed in a
single-layer graphene through the mechanism of magnetic catalysis of
dynamical mass generation as pointed out in
\onlinecite{Gusynin_PRL}. This work showed that the magnetic
catalysis can induce exciton condensation even for weak
particle-hole coupling\cite{Gusynin_PRB}. These results are obtained
in the framework of quantum electrodynamics $QED$ deduced from the
linear energy spectrum of the graphene monolayer.

In the case of a bilayer, additional excitonic channels become
possible as the excess electrons and holes from the two layers can
form a ``real-space" exciton. In this paper, we consider the
possibility of an excitonic instability in the biased graphene
bilayer in the framework of Hartree-Fock theory. A conventional
Hartree-Fock treatment had been used in the past to understand the
exciton formation in semiconductors with
success\cite{halperin-rice-review}. It is shown that the exciton can
be formed if the strength of the Coulomb interaction $U$ is larger
than a threshold value $U_c$ which, for realistic graphene
parameters, is comparable to the intra-layer hopping energy. The
threshold $U_c$ is, in turn, bias dependent and can be tuned to a
minimum value for an optimal bias $V_{o}$. Moreover, a reduction of
the inter-layer hopping perhaps through intercalation is shown to
greatly reduce the threshold value $U_c$.

In identifying excitonic channels, we consider two possible
scenarios. One is the pairing through the shortest-distance
neighbors between the layers ($a-d$ dimer in Fig.
\ref{fig1:GrapheneBilayer}), and the other, through the second
shortest-distance neighbors between the layers ($a-c$ and $b-d$
dimers in Fig. \ref{fig1:GrapheneBilayer}). For each scenario we
identify the threshold interaction strength $U_c$ and its dependence
on the bias and the inter-layer hopping parameter.

This work is divided into the following sequence. Section
\ref{Section2} describes the graphene bilayer and its model
Hamiltonian including the short-range Coulomb interaction across the
layers. Two excitonic channels we will consider in this paper are
introduced. In the following two sections, each of these
possibilities are examined in detail using the appropriate gap
equations and their solutions. The work is summarized in section
\ref{Section5}. Some of the technical aspects are summarized in the
Appendix.

\section{Formulation of the Exciton Problem\label{Section2}}

\begin{figure}
\epsfig{file=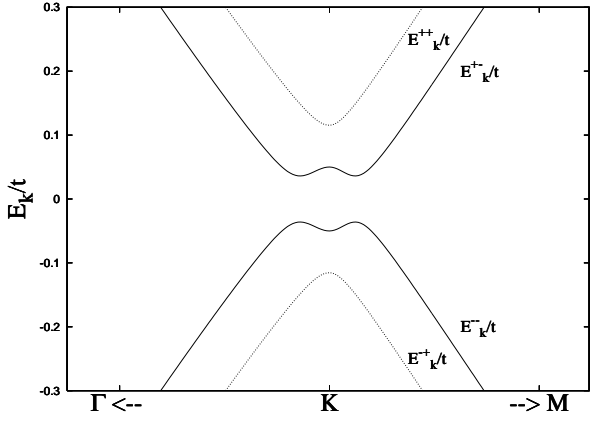,width=8cm}
\caption{Energy spectrum
for the graphene bilayer with $t=2.9$ eV, $t_\perp/t = 0.052$ and
$V/t = 0.05$. $E^{\pm -}_k/t$ in full line and $E^{\pm +}_k/t$ in
dashed line. See text for definition of the energy branches labeled
by $E^{\pm \pm}_k$. } \label{fig3:Spectrum}
\end{figure}

\begin{figure}
\epsfig{file=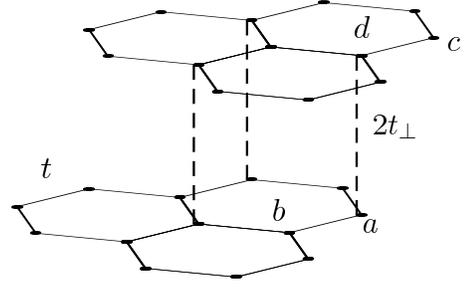,width=6cm}
\caption{Graphene bilayer (Bernal stacking).
The $a$-$d$ dimer is depicted as dashed lines.}
\label{fig1:GrapheneBilayer}
\end{figure}

Graphene bilayer is a two honeycomb array stacked in a Bernal
arrangement as depicted in Fig. \ref{fig1:GrapheneBilayer}. In each
layer the electrons can hop between nearest-neigbhour carbon atoms
through $\pi$-orbitals with energy $t$, which is typically assumed
at 2.9 eV. In a Bernal stacking electrons are allowed to do
inter-layer hopping through the $a$-$d$ dimers with the hopping
energy given as $2 t_\perp$ with the $t_\perp/t$ approximately
0.052. Here a dimer is defined as the pair of carbon atoms from the
adjacent layers stacked along the $c$-axis.

In writing down the Hamiltonian appropriate for the graphene
bilayer, we denote the electron operators for the two sublattices in
the lower layer by $a_i$ and $b_i$, and those in the upper layer by
$c_i$ and $d_i$. We assume a symmetric doping due to the external
bias $\pm V$ with excess electrons and holes on the lower and upper
graphene layers, respectively. We will be interested in the
formation of the same-spin electron-hole exciton here, hence the
spin degree of freedom $\sigma$ will be dropped. The Hamiltonian of
the graphene bilayer in the absence of the Coulomb interaction reads

\ba H_0 && =  -t \sum_{\ij} \left(a^\dagger_{i} b_{j} + b_{j}^\dagger
a_{i}  + c^\dagger_{i} d_{j} + d_{j}^\dagger c_{i}\right) \nn
&& -2 t_\bot \underset{i}{\sum} \left(a^\dagger_{i} d_{i} +
d^\dagger_{i} a_{i} \right) \nn
&& + V\underset{i}{\sum} \left(c^\dagger_{i} c_{i} + d^\dagger_{i}
d_{i}  -a^\dagger_{i} a_{i} - b^\dagger_{i} b_{i} \right) .
\label{BilayerHamiltonian} \ea
After the diagonalization (Derivation is given in Appendix
\ref{AppendixA}), Eq. (\ref{BilayerHamiltonian}) is transformed to

\ba
H_0 &=& \underset{k}{\sum} \left(
\begin{array}{cccc}
\alpha_k^\dagger &
\beta_k^\dagger &
\gamma_k^\dagger &
\delta_k^\dagger
\end{array}
\right)
\notag \\
&& \times
\left(\begin{array}{cccc}
E^{++}_k & 0 & 0 & 0 \\
0 &  E^{+-}_k & 0 & 0 \\
0 & 0 &  E^{--}_k & 0\\
0 & 0 & 0 & E^{-+}_k
\end{array}
\right)
\left(
\begin{array}{c}
\alpha_k \\
\beta_k \\
\gamma_k \\
\delta_k
\end{array}
\right), \label{diagonalized-H}
\ea
where $(\alpha_k, \beta_k, \gamma_k, \delta_k)$ now serve to define
the eigenstates. The energy spectra depicted in Fig.
\ref{fig3:Spectrum} are the ones given by

\be E^{\pm \pm}_k = \pm \sqrt{{\varepsilon_k}^2 + 2 t_\perp^2  + V^2
\pm 2 \sqrt{ t_\perp^4 + {\varepsilon_k}^2 ( t_\perp^2 + V^2 )}}.
\ee
The bare kinetic energy $\varepsilon_k$ within the monolayer reads
$\varepsilon_k = t|\sum_{\alpha=1}^3  e^{i k.e_\alpha}|$, where
$e_\alpha$'s are the nearest-neighbor vectors of the graphene
monolayer: $e_1 = (1,0)$, $e_2 = (-1/2,\sqrt{3}/2) $ and $e_3 =
(-1/2,-\sqrt{3}/2)$.

The two independent nodal points $K_{1(2)}$ where the bare electron
spectrum $\varepsilon_k$ vanishes are chosen as
$K_1 = \left(0,\frac{4\pi}{3\sqrt{3}}\right)$, $K_2 = -K_1 $ in the basis
$(e_x,e_y)$ in the Brillouin zone.
The sum $\sum_\alpha e^{i k.e_\alpha}$ is approximately given by $-(3/2)
(k_y \!-\! i k_x)$ near $K_1$ and $(3/2) (k_y \!+\! i k_x) $ near
$K_2$.

The bottom of the lower conduction band, $E_k^{+-}$, occurs at $k$
points where $\varepsilon_k^2 = (\varepsilon_k^2 )_m = V^2 (V^2
+2t_\bot^2 ) /(V^2 + t_\bot^2 )$, with the energy $E_m = t_\bot
V/\sqrt{t_\bot^2 + V^2}$. The energy gap separating the valence and
conduction bands is twice this value. The energy difference between
the two conduction bands or the two valence bands is $\sqrt{V^2
+2t_\bot^2} - V$ at $\varepsilon_k = 0$, and $\sqrt{[ 4(V^4 +
t_\bot^4 )+9 V^2 t_\bot^2 ]/(V^2 + t_\bot^2) }- V t_\bot /\sqrt{V^2
+t_\bot^2 }$ at $\varepsilon_k = ( \varepsilon_k )_m$. These two
quantities approach $t_\bot^2 /V$ and $2V$, respectively, as
$V/t_\perp \rightarrow \infty$. Generally, the presence of both
inter-layer hopping and the bias is essential in producing the gaps
separating the various bands as depicted in Fig.
\ref{fig3:Spectrum}.

In describing the exciton formation, we propose to use the
inter-layer interaction truncated to the second nearest neighbors as

\ba V_C &=& U_1 \sum_i n_{a,i}n_{d,i} \nn &&+ U_2 \sum_{i \alpha}
\left( n_{a,i} n_{c,i-e_\alpha}+n_{b,i}n_{d,i-e_\alpha} \right)
.\label{eq:inter-layer-Coulomb}\ea
The local electronic densities are given by $n_{a,i} = \abar_i a_i$,
etc. The total Hamiltonian then reads $H = H_0 + V_C$. The $U_1$ and
$U_2$ terms are responsible for the exciton formation across the
$a-d$ dimer (nearest neighbor) and the $a-c$, $b-d$ dimer (second
nearest neighbor), respectively.

At this point, several mean-field decoupling strategies present
themselves. The average $\langle \abar_i d_i \rangle$ might be a
candidate order parameter for the exciton pairing, but this quantity
is nonzero even in the absence of any inter-layer interaction,
provided the inter-layer tunneling $t_\perp$ remains non-zero. Only
when $t_\perp = 0$ does this average become the exact order
parameter. Nevertheless, one can use the ``difference" (to be
quantified in the next section) of $\langle \abar_i d_i\rangle$
obtained in the presence and absence of excitons as the order
parameter. This is the strategy we adopt to discuss the $a-d$ dimer
exciton formation.

For the second-neighbor interaction, we could think of averages like
$ \langle \abar_i c_{i-e_\alpha} \rangle$, and $\langle \bbar_i
d_{i-e_\alpha}\rangle $, as possible excitonic order parameters.
Again, these averages are non-zero even in the absence of the
interaction $V_C$. However, since averages $\langle \abar_i
c_{i-e_\alpha}\rangle $ for $\alpha=1,2,3$ are related by the $Z_3$
symmetry, one could form linear combinations $\sum_\alpha u_\alpha
\langle \abar_i c_{i-e_\alpha} \rangle$ which remains zero in the
non-interacting case, but becomes a nonzero value once the
interaction $U_2$ is turned on and excitons are formed. The
appropriate linear combination is easily identified. For the second
nearest-neighbor pairing, the excitonic order is directly related to
the loss of $Z_3$ rotational symmetry of the lattice.

Finally, we assume that at low energy the main mechanism of the
exciton formation is due to the hybridization of the upper valence
band $(E^{+-}_k)$ and lower conduction bands ($E^{--}_k$), while the
two outlying ones, $E^{++}_k$ and $E^{-+}_k$, remain as spectators.
Accordingly the following reduced Hamiltonian may be used instead of
Eq. (\ref{diagonalized-H}):

\be H' = \sum_k E_k (\betabar_k \beta_k - \gammabar_k \gamma_k ), ~~
E_k = E^{+-}_k  = - E^{--}_k . \ee
Note that the two outlying bands are separated from the two inner
ones by an energy difference that grows as $V$ when $V/t_\perp$ is
sufficiently large. The truncation scheme is expected to be valid
when the bias $V$ far exceeds the inter-layer tunneling energy; a
situation easily realized in tunable gate
systems\cite{Castro,OOstinga}. The inter-layer interaction, Eq.
(\ref{eq:inter-layer-Coulomb}), will be truncated in the same
subspace spanned by $(\beta_k, \gamma_k )$. Such truncation greatly
simplify the algebra in subsequent discussions.

\section{First Neighbor Exciton Pairing \label{Section3}}

The first-neighbor interaction part reads

\be V_{C} (U_1 ) = U_1 \sum_i n_{a,i} n_{d,i} = U_1 \sum_{q k k'}
a_{k+q}^\dagger a_{k} d_{k' -q}^\dagger d_{k^{'}}. \label{V_C} \ee
According to our truncation scheme, the various operators can be
expanded in terms of $\beta_k$ and $\gamma_k$ operators
corresponding to the lower conduction and upper valence bands,
respectively.

\begin{eqnarray*}
a_k &=& U_{12}(k) \beta_k + U_{13}(k) \gamma_k,  \nn d_k &=&
U_{42}(k) \beta_k + U_{43}(k) \gamma_k .
\end{eqnarray*}
The $4\times4$ unitary matrix $U$ diagonalizing the Hamiltonian
\eqref{BilayerHamiltonian} (see appendix \ref{AppendixA} for a
description of $U$) is used. The first-neighbor Coulomb interaction
in the truncated space reads

\ba V_{C} (U_1) &=& U_1 \underset{qkk'}{\sum} \left( U^{*}_{12}(k+q)
\betabar_{k+q} + U^{*}_{13}(k+q) \gammabar_{k+q} \right)  \times \nn
&& ~~~~~\left( U_{12}(k) \beta_{k} + U_{13}(k) \gamma_{k}
\right)\times \nn &&  \left( U^{*}_{42}(k' \!-\! q) \betabar_{k' -q}
+ U^{*}_{43}(k' \!-\! q) \gammabar_{k' -q} \right)\times  \nn &&
~~~~~\left( U_{42}(k' ) \beta_{k'} + U_{43}(k') \gamma_{k'} \right).
\label{full-truncated-Coulomb} \ea
As our main concern is to explore the possibility of the excitonic
order represented by nonzero $\langle \gamma^\dagger_k \beta_k
\rangle$, we will only keep terms from Eq.
(\ref{full-truncated-Coulomb}) involving an even number of $\beta$
and $\gamma$ operators. In a Hartree-Fock approximation the
mean-field Hamiltonian using the exciton order parameter for the
$\beta$-$\gamma$ hybridization can be written down as

\be V_{C} (U_1 ) = - \sum_k ( \Delta_k \betabar_k \gamma_k +
\Deltabar_k \gammabar_k \beta_k ) .\label{CoulombInteraction} \ee
The exciton gap $\Delta_k$ is related to the exciton order parameter
$\langle \gamma^\dagger_k \beta_k \rangle$ through

\ba && \Delta_k = U_1 \left( \sum_q U_{12}(q) U^{*}_{43}(q) \langle
\gammabar_{q}\beta_{q}  \rangle\right) U^{*}_{12}(k)  U_{43} (k) \nn
&& + U_1 \left( \sum_q U^{*}_{13} (q) U_{42} (q) \langle
\gammabar_{q} \beta_{q}\rangle \right) U_{13} (k) U^{*}_{42} (k) \nn
&& - U_1 \left( \sum_q U_{42} (q) U^{*}_{43} (q)  \langle
\gammabar_{q} \beta_{q} \rangle \right) U^{*}_{12} (k) U_{13} (k)
\nn && - U_1  \left( \sum_q U_{12} (q) U^{*}_{13} (q)  \langle
\gammabar_{q} \beta_{q} \rangle \right) U^{*}_{42}(k) U_{43}(k) .
\label{gap-equation} \ea
Combining the kinetic part and the mean-field Coulomb interaction
$V_C$ one obtains the full Hamiltonian

\ba
H = \sum_k E_k (\betabar_k \beta_k - \gammabar_k \gamma_k )
-\sum_k ( \Delta_k \betabar_k \gamma_k + \Deltabar_k \gammabar_k
\beta_k ).
\notag \\ \label{Eq11}
\ea
The Hamiltonian \eqref{Eq11} can be further diagonalized by the $2\times2$
unitary rotation

\be
\left(
\begin{array}{c} \beta_k \\ \gamma_k \end{array}\right) = \left(
\begin{array}{cc} e^{iy_k} \cos \theta_{5k} & e^{iy_k} \sin \theta_{5k} \\
-\sin \theta_{5k}  & \cos \theta_{5k} \end{array}\right)\left(
\begin{array}{c} B_k \\ C_k \end{array}\right),
\ee
with $e^{iy_k} = \Delta_k / |\Delta_k | $, and $\cos 2 \theta_{5k} =
E_k /\mathcal{E}_k $, $\sin 2\theta_{5k}  = |\Delta_k |
/\mathcal{E}_k$. In terms of the eigen-operators $B_k$, $C_k$ and
the eigenvalue $\mathcal{E}_k =\sqrt{E_k^2 + |\Delta_k|^2 }$, the
Hamiltonian \eqref{Eq11} reads $H= \sum_k \mathcal{E}_k (\Bbar_k B_k
- \Cbar_k C_k )$. The hybridization is given by

\be
\langle \gammabar_k \beta_k \rangle  = \frac{\Delta_k}{2\mathcal{E}_k }
\tanh \left(\frac{\beta \mathcal{E}_k} { 2} \right).
\label{gap-eq-1}
\ee
By inserting expressions of the unitary matrix elements
(\ref{expression-for-U}) and the hybridization
(\ref{gap-eq-1}) in Eq. (\ref{gap-equation}), one readily
finds that the phase of the gap function is dictated in the
manner

\begin{eqnarray}
\Delta_k = e^{-i\phi_k} |\Delta_k |,~~e^{i\phi_k} =
\frac{\sum_\alpha e^{ik\cdot e_\alpha}}{|\sum_\alpha e^{ik\cdot
e_\alpha}|} . \label{eq:phase-in-Delta}\end{eqnarray}
The phase factor in the excitonic gap has a winding of $2\pi$ around
one Dirac point, and $-2\pi$ around the other. The total winding
around the circumference of the full Brillouin zone is therefore
zero.

A general connection between the phase singularity of the wave
function and the singularities in the order parameters was
considered in Ref. \onlinecite{Murakami}. This discussion can also
be applied to excitonic order. In two dimensions, the topological
structure discussed in Ref. \onlinecite{Murakami} is defined as the
total number of phase winding for the whole Brillouin zone, which in
this case is zero. In fact, the phase winding around $K_1$ and $K_2$
in Eq. (\ref{eq:phase-in-Delta}) can be removed by a gauge
transformation\cite{thanks}

\begin{equation}
\beta'_k=\beta_k e^{i\phi_{k}},\
 \gamma'_k=\gamma_k.
\end{equation}
With this transformation,
$\langle\gamma'^{\dagger}_{k}\beta'_{k}\rangle$ becomes real, and
the phase vanishes at $K_1$ and $K_2$.

\begin{figure}
\epsfig{file=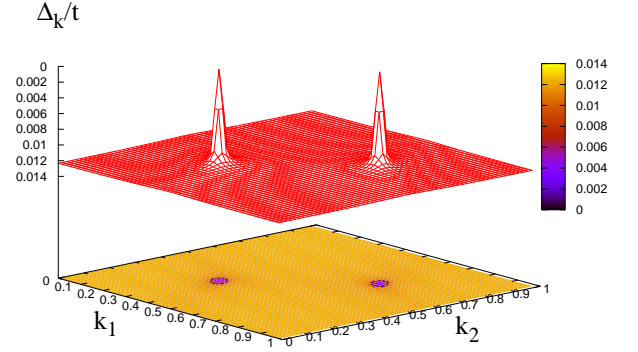,width=8.5cm}
\caption{(color online) Exciton gap obtained from equation
\eqref{gap-equation-reduced} at $T=0$. Here the parameters are : $t
= 2.9$ eV, $V/t = 0.1$, $U_1 /t = 8.5$, $t_\perp/t = 0.052$ for a
$60 \times 60$ lattice in the reciprocal space spanned by $k = k_1
\textbf{R}_1 + k_2 \textbf{R}_2$. The reciprocal vectors
$\textbf{R}_1$ and $\textbf{R}_2$ are defined by $\textbf{R}_1 =
\frac{2\pi}{3}(1,\sqrt{3})$ and $\textbf{R}_2 =
\frac{2\pi}{3}(-1,\sqrt{3})$.} \label{Exciton_Gap_N60}
\end{figure}

Taking out the phase, the gap equation (\ref{gap-equation}) becomes

\ba && | \Delta_k |
 = \frac{U_1 }{16} \sum_p (1\!-\!\sin 2\theta_{2p})(1\!-\!\sin
2\theta_{2k}) \nn && \times [ 1\! +\!\cos (2\theta_{4p} \!
-\!2\theta_{4k} )] \frac{|\Delta_p |}{\mathcal{E}_p}\tanh \left(
\frac{\beta \mathcal{E}_p}{2} \right). \label{gap-equation-reduced}
\ea
The various factors are defined in Appendix \ref{AppendixA}. The
equation can be solved numerically for given values of $U$, $V$ and
the inter-layer hopping parameter $t_\perp$. The numerical solution
of Eq. \eqref{gap-equation-reduced} is depicted in Fig.
\ref{Exciton_Gap_N60}. Regarding the momentum dependence of the
exciton gap we see that it increases from zero at the nodal points
to reach saturation far from $K_1$ and $K_2$.

\begin{figure}
\epsfig{file=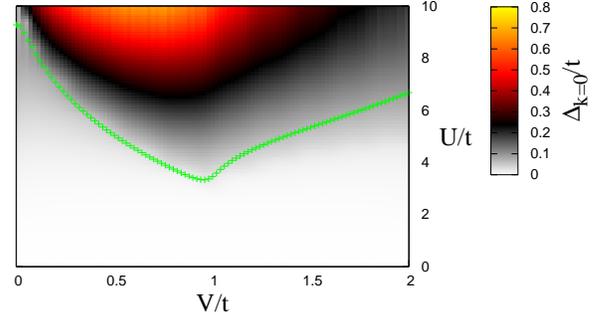,width=8cm}
\caption{(color online) Zero temperature exciton gap magnitude
$|\Delta_k |$ at $k=0$ (half way between $K_1$ and $K_2$) depending
on the short-ranged Coulomb interaction $U_1$ and the bias $V$
applied on the layers, obtained for $30\times30$ lattice with $t =
2.9$ eV, $t_\perp/t=0.052$. The green line correspond to the
threshold $U_{1c} (V)$ solution of Eq. (\ref{gap-equation-reduced})
for a lattice size of $500 \times 500$ carbon atoms.}
\label{Gap_vs_U_V}
\end{figure}

Figure \ref{Gap_vs_U_V} shows the solution of Eq.
\eqref{gap-equation-reduced} for a range of $U_1$ and various values
of the bias $V$. At zero temperature a second order phase transition
of the exciton gap takes place with respect to the Coulomb
interaction $U_1$ for each given bias $V$. This threshold value
$U_{1c}$ at which excitons begin to form is a function of $V$ and is
shown as a green line in Fig. \ref{Gap_vs_U_V}. $U_{1c} (V)$ reaches
a minimal value $U_{1c}/t \simeq 3.5$ for an ``optimal" choice of
the bias $V_{1o}$ which is found at $V_{1o}/t \simeq 1$.

\begin{figure}
\epsfig{file=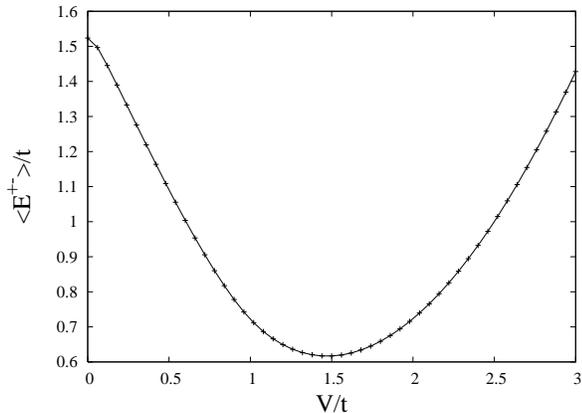,width=8cm} \caption{Energy average $
\langle E^{+-}_k \rangle$ depending on the bias $V$. The average of
the energy is computed by summing $E^{+-}_k$ over the whole
Brillouin zone with $t=2.9$ eV and $t_\bot /t = 0.052$.}
\label{EnergyAverage}
\end{figure}

Interestingly, the dependence on the bias $U_{1c} (V)$ appears to be
related to the behavior of the exciton gap $\Delta_{k=0}$, obtained
far away from the Dirac points at $k=0$, as shown in Fig.
\ref{Gap_vs_U_V}. The non-monotonic dependence of the gap value on
$V$ is apparent. A similar behavior is observed in the
conduction-valence band energy gap, as exemplified in the Brillouin
zone average $\langle E^{+-}_k\rangle = \sum_{k \in BZ} E^{+-}_k $
shown in Fig. \ref{EnergyAverage}.

\begin{figure}
\epsfig{file=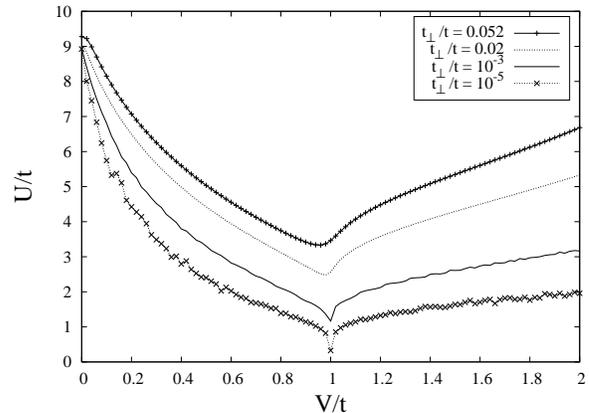,width=8cm} \caption{Threshold $U_{1c}$
depending on the bias $V$ at zero temperature, $t=2.9$ eV and
various values of $t_\perp$.} \label{Uc_V_t}
\end{figure}

Using Eq. (\ref{gap-equation-reduced}) we can deduce the dependence
of $U_c (V)$ on the inter-layer parameter $t_\perp$.  As Fig.
\ref{Uc_V_t} shows, the threshold value decreases with $t_\perp$ and
tends to zero as $t_\perp /t \rightarrow 0$. Reducing the
inter-layer hopping parameter would reduce the threshold $U_{1c}$ of
the short-ranged Coulomb interaction above which excitons can form.
Intercalation of layers of non-doping and insulating atoms between
the two carbon layers would reduce significantly the inter-layer
hopping parameter toward zero. The concomitant reduction in the
Coulomb interaction $U_1$ with distance will be sufficiently slow
compared to the exponential decay of $t_\perp$, so that the regime
$U>U_c (V)$ can be attained for a range of bias around $V_{1o}$. Our
analysis suggests that searching for ways to reduce the inter-layer
hopping parameter experimentally would shed more light on the
physics of exciton formation in graphene bilayer.

\begin{figure}
\epsfig{file=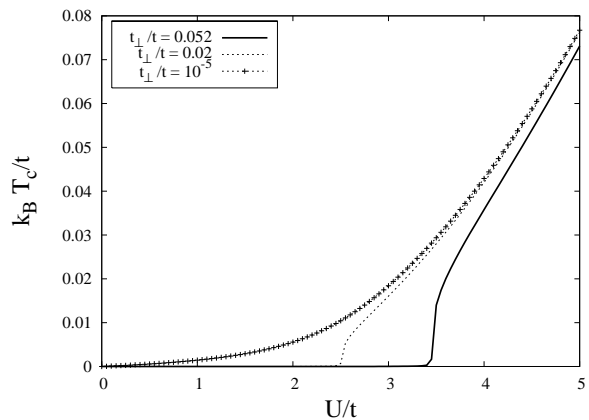,width=8cm} \caption{Critical temperature
$T_{c}$ depending on the ratio $U_1 /t$ for several values of
$t_\perp /t$.  The critical temperature has been computed for $100
\times 100$ sublattice size and for the optimal value of the bias
$V/t = 1$. $T_c$ tends to zero when $t_\perp /t \rightarrow 0$. }
\label{Tc_U}
\end{figure}

Finally, Fig. \ref{Tc_U} depicts the behaviour of the critical
temperature $T_c$ with respect to the Coulomb interaction $U$ and
for various hopping parameters $t_\perp$.

\section{Second Neighbor Exciton Pairing \label{Section4}}

Our  approach in the previous section was based on the interaction
with $U_2 = 0 $. The average $\langle \abar_i d_i \rangle$ read

\ba &&\langle \abar_i d_i \rangle = \sum_k \langle \abar_k d_k
\rangle \nn
&& = \sum_k \left(U^*_{12} (k) U_{42} (k) \langle \betabar_k \beta_k
\rangle + \Ubar_{13} (k) U_{43}(k) \langle \gammabar_k \gamma_k
\rangle \right) \nn
&& + \sum_k \left( \Ubar_{12} (k) U_{43}(k) \langle \betabar_k
\gamma_k \rangle + \Ubar_{13}(k) U_{42}(k) \langle \gammabar_k
\beta_k \rangle \right)\nn
\ea
in the scheme where the lowest and highest energy bands were
truncated out. This average is nonzero even without the excitons at
arbitrary temperature, and is not a good measure of the possible
phase transition in the model. Instead, we relied on the fact that
$\langle \betabar_k \gamma_k\rangle$ is zero unless the excitons
exist, and used this average as a measure of the excitonic order and
excitonic phase transition in the model. And indeed this order
parameter vanished at high enough temperature and/or weak enough
coupling, allowing us to identify the critical points, and so forth.

In this section, we search for an excitonic order parameter defined
in real space, which also vanishes identically for a non-excitonic
phase. The averages $\langle \cbar_{i-e_\alpha} a_i\rangle$ and
$\langle \dbar_{i-e_\alpha} b_i \rangle $ are given by

\ba && \langle\cbar_{i-e_\alpha} a_i \rangle = \sum_k e^{ik\cdot
e_\alpha} e^{-i\phi_k} f(k),  \nn
&& \langle\dbar_{i-e_\alpha} b_i \rangle = - \sum_k e^{ik\cdot
e_\alpha} e^{-i\phi_k} f(k),
\label{eq:possible-OP}\ea
in the non-interacting limit, Eq. (\ref{diagonalized-H}). The three
unit vectors $e_\alpha$ were defined earlier as the difference of
the nearest neighbor positions in a given graphene layer. Here
$f(k)$ is a function whose detailed form is unimportant to us. The
combination $e^{-i\phi_k} f(k)$ is symmetric under the 120$^\circ$
rotation of the $k$ vector, which in turn implies that $\langle
\cbar_{i\!-\!e_1} a_i \rangle = \langle \cbar_{i\!-\!e_2}a_i \rangle
= \langle \cbar_{i\!-\!e_3}a_i \rangle$, and $\langle
\dbar_{i\!-\!e_1} b_i \rangle = \langle \dbar_{i\!-\!e_2}b_i \rangle
= \langle \dbar_{i\!-\!e_3}b_i \rangle$.

This observation suggests a strategy for defining an appropriate
order parameter. First define $h_{i\alpha} = \cbar_{i-e_\alpha} a_i$
and $g_{i\alpha} = \dbar_{i-e_\alpha} b_i$, then one can form the
following linear combinations
\ba \chi_i^{(1)} = h_{i1} -\frac{1}{2}\left( h_{i2} +  h_{i3}
\right), && \Xi_i^{(1)} = g_{i1} -\frac{1}{2}\left( g_{i2} +  g_{i3}
\right) \nn
\chi^{(2)}_i = h_{i2} -\frac{1}{2}\left( h_{i1} + h_{i3} \right), &&
\Xi^{(2)}_i = g_{i2} -\frac{1}{2}\left( g_{i1} + g_{i3} \right) \nn
\chi^{(3)} = h_{i3} -\frac{1}{2}\left( h_{i1} + h_{i2} \right), &&
\Xi^{(3)}_i = g_{i3} -\frac{1}{2}\left( g_{i1} + g_{i2} \right).
\nn\ea The operators $\chi^{(\beta)}_i$ and $\Xi^{(\beta)}_i$ have a
zero average value in the non-excitonic phase, $U_2=0$, due to the
underlying $Z_3$ symmetry. In turn, non-zero value of one of the
averages implies the $Z_3$ symmetry is spontaneously broken.

The short-ranged Coulomb interaction \eqref{eq:inter-layer-Coulomb}
with $U_1=0$ and $U_2 \neq 0$ will render the mean-field Hamiltonian

\ba && -U_2 \sum_{i,\alpha} \left( \langle \abar_i c_{i-e_\alpha}
\rangle \cbar_{i-e_\alpha} a_i + h.c.\right)\nn
&& -U_2 \sum_{i,\alpha} \left( \langle \bbar_i d_{i-e_\alpha}\rangle
\dbar_{i-e_\alpha} b_i +h.c.\right).
\label{Eq15}
\ea
In terms of the new operator $\chi^{(\beta)}_i$ and
$\Xi^{(\beta)}_i$ just defined, it can be recast in the form

\begin{eqnarray}
&&V_C(U_2) = -\frac{4}{9} U_2 \sum_{i}
\Bigg\{
\sum_\alpha \left( \langle {\chi^{(\alpha)}_i}^\dagger \rangle
\chi^{(\alpha)}_i + \langle {\Xi^{(\alpha)}_i}^\dagger \rangle
\Xi^{(\alpha)}_i \right) \nn
&& ~~~~~+\frac{3}{4} \langle h^\dagger_{i,1} + h^\dagger_{i,2} +
h^\dagger_{i,3} \rangle \left( h_{i,1} + h_{i,2} + h_{i,3} \right)
\nn && ~~~~~~ +  ( h_{i\alpha} \rightarrow g_{i\alpha}  )+ h.c.
\Bigg\}. \label{Eq17}
\end{eqnarray}
The Coulomb interaction expressed in Eq. \eqref{Eq17} is fully $Z_3$
symmetric (see Appendix \ref{AppendixB} for the full expression of
Eq.  \eqref{Eq17} in terms of the operator $h_i$ only). We remark
that the second line of Eq. \eqref{Eq17} is irrelevant for the
exciton formation and can be dropped.

Assuming translational invariance we can take $\langle
\chi^{(\beta)}_i\rangle = \langle \chi^{(\beta)}\rangle$ and
$\langle \Xi^{(\beta)}_i \rangle=\langle \Xi^{(\beta)} \rangle  $,
and express the interaction as

\ba V_C(U_2) &=& - \sum_k \left( \Theta_k \beta_k^\dagger \gamma_k +
\Theta^*_k \gamma_k^\dagger \beta_k \right) \ea
where $\Theta_k$ expresses the exciton gap. Using the total
Hamiltonian  $H = \sum_k E_k (\beta^\dagger_k \beta_k -
\gamma^\dagger_k \gamma_k ) -\sum_k \Big( \Theta_k \beta^\dagger_k
\gamma_k + \Theta^*_k \gamma^\dagger_k \beta_k \Big)$, one can
derive the averages $\langle \chi^{(\beta)} \rangle$ and $\langle
\Xi^{(\beta)}\rangle$ of the exciton order parameter from the
self-consistent equations

\begin{eqnarray}
\langle \chi^{(\beta)} \rangle &=& \frac{U_2}{9} \sum_k
\widetilde{\varphi}_k^{(\beta)} \cos^2(2\theta_{2k})
\frac{1}{\mathcal{E}_k}\tanh \frac{\beta \mathcal{E}_k}{2} \nn
&& \times \Bigg[ i \cos^2(2\theta_{1k}) \text{ Im }
\left(\sum_{\beta^{'}} \varphi_k^{(\beta^{'})} \langle
\chi^{(\beta^{'})} \rangle \right) \nn &&+ \cos^2(2\theta_{4k})
\text{ Re } \left(\sum_{\beta^{'}} \varphi_k^{(\beta^{'})} \langle
\chi^{(\beta^{'})} \rangle \right) \Bigg]
\notag \\
\label{Eq18}
\end{eqnarray}
for $\beta=\{1,2,3\}$. The averages $\langle \Xi^{(\beta)} \rangle$
are related to $\langle \chi^{(\beta)} \rangle$ by the simple
relation $\langle \Xi^{(\beta)} \rangle = -\langle \chi^{(\beta)}
\rangle$ for any $\beta$. We defined

\ba  \varphi_k^{(1)} &=& e^{i\phi_k} \left[e^{ik.e_1}
-\frac{1}{2}\left(e^{ik.e_2} +e^{ik.e_3} \right) \right] \nn
\widetilde{\varphi}_k^{(1)} &=& e^{-i\phi_k} \left[e^{ik.e_1}
-\frac{1}{2}\left(e^{ik.e_2}+e^{ik.e_3} \right) \right] \ea
and the $Z_3$ symmetric counterparts $\varphi_k^{(2)}$,
$\varphi_k^{(3)}$, $\widetilde{\varphi}_k^{(2)}$, and
$\widetilde{\varphi}_k^{(3)}$ accordingly. The energy of the
quasi-particles reads $\mathcal{E}_k = \sqrt{E_k^2 + |\Theta_k |^2}$
and the exciton gap is given by

\begin{eqnarray}
&&  ~~~~~~ e^{i\phi_k} \Theta_k = \nn
&& \frac{U_2}{9} \cos{(2\theta_{2k})} \Bigg[
\cos{(2\theta_{1k})}\times 4 i \text{ Im } \left(\sum_{\beta}
\varphi_k^{(\beta )} \langle \chi^{\beta} \rangle \right) \nn
&& + \cos{(2\theta_{4k})} \times 4  \text{ Re } \left(\sum_{\beta}
\varphi_k^{(\beta)} \langle \chi^{\beta} \rangle \right) \Bigg]
.\label{Eq24}
\end{eqnarray}

The system of self-consistent equations \eqref{Eq18} admits an
ensemble of solutions all obeying $\sum_\beta \langle \chi^{(\beta)}
\rangle = 0$. As it turns out, the numerical solution always follows
the condition that two of the $| \chi^{(\alpha)} |$'s are the same
and different from the third. Furthermore, the phases of the two
equal-amplitude bonds can be made equal through phase re-definition
of the operators, and we can choose, for instance, $\chi^{(2)}
=\chi^{(3)} \neq \chi^{(1)}$ without loss of generality. The other
choices are related by $Z_3$ permutation.

We will now exclusively consider the configuration $\langle
\chi^{(1)} \rangle \neq \langle \chi^{(2)} \rangle = \langle
\chi^{(3)} \rangle$ where, due to $\sum_\beta \chi^{(\beta)} =0$,
the following relation holds:

\begin{eqnarray}
\langle \chi^{(1)} \rangle = -2 \langle \chi^{(2)} \rangle = -2
\langle \chi^{(3)} \rangle .\label{Eq4}
\end{eqnarray}
Introducing relation \eqref{Eq4} into Eq. \eqref{Eq18}, one gets a
single self-consistent equation of the exciton instability

\begin{eqnarray}
&&\langle \chi^{(1)} \rangle = \frac{U_2}{9} \sum_k
\widetilde{\varphi}_k^{(1)} \cos^2(2\theta_{2k})
\frac{1}{\mathcal{E}_k}\tanh \frac{\beta \mathcal{E}_k}{2} \nn
&& \times \Bigg[ i \cos^2(2\theta_{1k}) \text{ Im } \left[
\left(\varphi_k^{(1)} -\frac{1}{2}\left( \varphi_k^{(2)} +
\varphi_k^{(3)} \right)\right) \langle \chi^{(1)} \rangle \right]
\notag \\
&& \quad + \cos^2(2\theta_{4k}) \text{ Re } \left[
\left(\varphi_k^{(1)} -\frac{1}{2}\left( \varphi_k^{(2)} +
\varphi_k^{(3)} \right)\right) \langle \chi^{(1)} \rangle \right]
\Bigg] . \nn \label{Eq5}
\end{eqnarray}
Solution of this can be used to generate the exciton gap $\Theta_k$
using Eq. (\ref{Eq24}).

\begin{figure}
\hspace*{-1cm} \epsfig{file=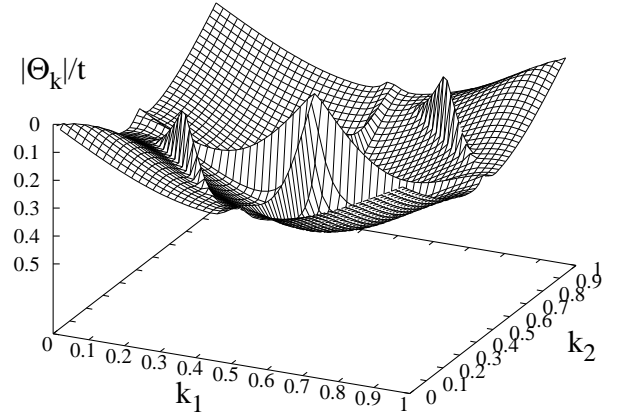,width=10cm}
\caption{The exciton gap amplitude $|\Theta_k|/t$ is plotted over
the whole Brillouin zone of the graphene bilayer. The parameters are
$t_\perp/t = 0.052$, $V/t = 1$. The Coulomb interaction $U_1 = 0$
and we have chosen an arbitrary value $U_2 /t = 3$. For a $50 \times
50$ lattice in the reciprocal space spanned by $k = k_1 \textbf{R}_1
+ k_2 \textbf{R}_2$.} \label{FigThetaQ}
\end{figure}

Figure \ref{FigThetaQ} represents the amplitude of the exciton gap
$|\Theta_k|$ over the whole Brillouin zone of the graphene bilayer.
The exciton gap vanishes at the Dirac nodal points $K_1$ and $K_2$
as well as for a wave vector $\mathbf{k} = \frac{1}{2}(\mathbf{R}_1
+ \mathbf{R}_2)$ where $\textbf{R}_1 = \frac{2\pi}{3}(1,\sqrt{3})$
and $\textbf{R}_2 = \frac{2\pi}{3}(-1,\sqrt{3})$. The vanishing of
the exciton amplitude at the point $\frac{1}{2}(\mathbf{R}_1 +
\mathbf{R}_2)$ marks the breaking of the $Z_3$ symmetry.

\begin{figure}
\center \rotatebox{-90}{
\epsfig{file=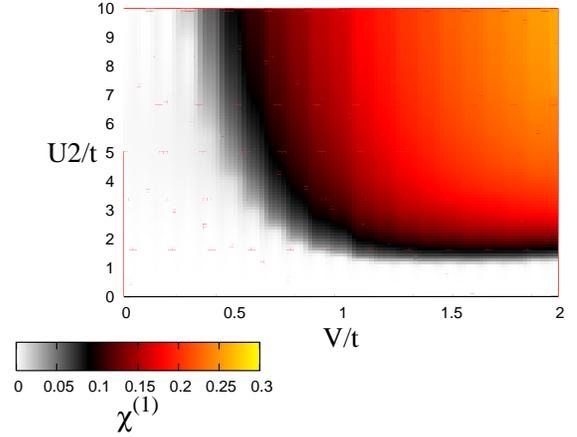,width=6cm}}
\caption{(color online) Contour of the amplitude of the exciton
average $\langle \chi^{(1)} \rangle$ in the configuration for which
$\langle \chi^{(1)} \rangle \neq \langle \chi^{(2)} \rangle \neq
\langle \chi^{(3)} \rangle$. Here $t = 2.9$ eV, $t_\perp /t =
0.052$, $V/t = 1$ and we used a sublattice of $30 \times 30$ carbon
atoms.} \label{Fig2}
\end{figure}

With Eq. (\ref{Eq5}), one can derive the threshold Coulomb
interaction strength which reads

\begin{eqnarray}
&&\frac{1}{U_{2c}} = \frac{1}{9} \sum_k \cos^2(2\theta_{2k})
\frac{1}{E_k}\tanh \frac{\beta E_k}{2} \nn
&& \times \Bigg[ \cos^2(2\theta_{4k}) \text{ Re }
\left(\widetilde{\varphi}_k^{(1)} \right) \text{ Re }
\left(\varphi_k^{(1)} -\frac{1}{2}\left( \varphi_k^{(2)} +
\varphi_k^{(3)} \right)\right) \nn
&& - \cos^2(2\theta_{1k}) \text{ Im }
\left(\widetilde{\varphi}_k^{(1)} \right) \text{ Im }
\left(\varphi_k^{(1)} -\frac{1}{2}\left( \varphi_k^{(2)} +
\varphi_k^{(3)} \right)\right) \Bigg].\nn
\end{eqnarray}

\begin{figure}
\center \epsfig{file=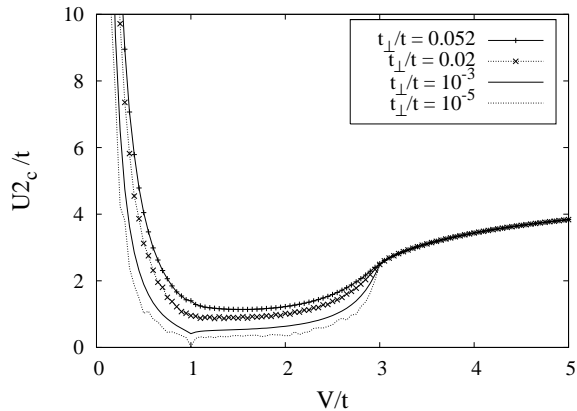,width=8cm}
\caption{Coulomb
threshold $U_{2c}$ depending on the bias $V$ for various inter-layer
hopping paramters $t_\perp$. We used $t = 2.9$ eV and sublattices of
$50 \times 50$ to $1200 \times 1200$ carbon atoms.} \label{Fig5}
\end{figure}

Figure \ref{Fig5} shows the variation of $U_{c2}$ with $V/t$ for
various values of $t_\perp /t$. The similarity of this plot to Fig.
\ref{Uc_V_t} is obvious. As for the case treating the Coulomb
interaction on dimer $a-d$ we see that there is an optimal value
$V_{2o}/t \simeq 1$ for which the threshold $U_{2c}$ is minimal.
Moreover as the inter-layer parameter $t_\perp$ is decreased (by
intercalation of insulating and non-doping atoms) the Coulomb
threshold decreases.

\begin{figure}
\center \epsfig{file=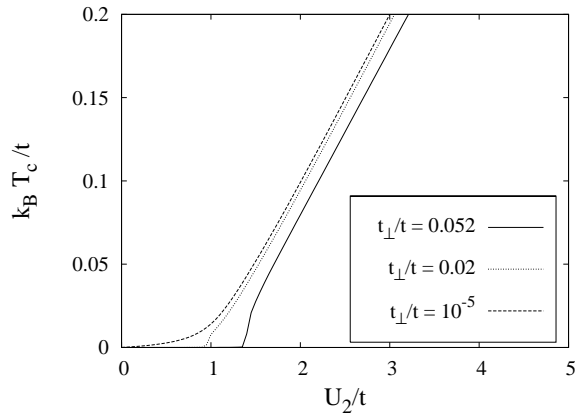,width=8cm}
\caption{Critical temperature for various Coulomb interaction and
different inter-layer hopping parameter $t_\perp$ as given in Eq.
Here $t = 2.9$ eV, $t_\perp/t = 0.052$, $V/t = 1$ and we used a
sublattice with $50 \times 50$ carbon atoms. Note the similarity to
$T_c$ plot in Fig. \ref{Tc_U}.} \label{Fig4}
\end{figure}

The critical temperature $T_c$ follows from

\begin{eqnarray}
&&1  = \frac{U_2}{9} \sum_k \cos^2(2\theta_{2k}) \frac{1}{E_k}\tanh
\frac{E_k}{2 k_B T_c} \nn
&& \times \Bigg[ \cos^2(2\theta_{4k}) \text{ Re }
\left(\widetilde{\varphi}_k^{(1)} \right) \text{ Re }
\left(\varphi_k^{(1)} -\frac{1}{2}\left( \varphi_k^{(2)} +
\varphi_k^{(3)} \right)\right) \nn
&& - \cos^2(2\theta_{1k}) \mathrm{ Im }
\left(\widetilde{\varphi}_k^{(1)} \right) \mathrm{ Im }
\left(\varphi_k^{(1)} -\frac{1}{2}\left( \varphi_k^{(2)} +
\varphi_k^{(3)} \right)\right) \Bigg]. \nn
\end{eqnarray}
Figure \ref{Fig4} depicts the variation of the critical temperature
with respect to the Coulomb interaction and for various values of
the inter-layer hopping parameter $t_\perp$. As for the dimer $a-d$
Coulomb interaction, exciton are formed at higher temperatures for
smaller $t_\perp /t$.

The behaviour observed in this section are in good agreement with
the behaviour of the critical temperature $T_c$ and the Coulomb
threshold $U_{1c}$ observed in the case treating the Coulomb
interaction $U_1$ on dimer $a-d$. However the Coulomb threshold
$U_{2c}$ is smaller than the threshold $U_{c1}$. For $t = 2.9$ eV,
$t_\perp /t = 0.052$, at the optimal value of the bias $V_o/t \simeq
1$, one gets $U_{1c}/t \simeq 3.5$, compared to $U_{2c}/t \simeq
1.5$.

\section{Conclusions \label{Section5}}

The graphene bilayer system was considered with a short-ranged
Coulomb interaction acting between the nearest and next-nearest
carbon sites in a Bernal stacking scheme of two carbon layers. The
short-ranged Coulomb interaction was introduced for both nearest
($U_1$ for $a-d$ dimer) and second-nearest ($U_2$ for $a-c$ and
$b-d$ dimers) neighbors between the two layers.

For a given bias $V$, or electron-hole imbalance between the layers,
a critical Coulomb interaction strength exists above which the
excitons form. For the first-neighbor, $a-d$ dimer interaction, the
critical strength is $U_c/t \simeq 3.5$ for a bias $V/t \simeq 1$.
The threshold becomes smaller in the case of only the second
neighbour Coulomb interaction and approximately equal to $U_{2c}/t
\simeq 1.5$ at $V/t \simeq 1$. Hence, doping by equal and opposite
charges of the bilayer system with the voltage difference applied
perpendicular to the bilayer can control the excitonic properties of
the graphene bilayer in a non-trivial way. The optimal value of the
bias $V$ (which gives rise to the least threshold value $U_c$) was
found to be $V_{o}/t \simeq 1$. This non-monotonic dependence on the
bias reflects the dependence of the energy gap between the
conduction and valence bands graphene bilayer on the same quantity.

Moreover, we showed that reducing the inter-layer hopping parameter,
$t_\perp \rightarrow 0$ reduces the threshold near the optimal bias
$U_c(V_o )$ to zero. We suggest that intercalation of non-doping and
insulating atomic layers between the carbon layers could reduce
significantly $t_\perp$ in such a way that the screened Coulomb
interaction $U$ obeys the condition $U>U_c$ (for bias around the
optimal value $V_{o}$) and excitons could form. It thus seems
possible that the formation of the exciton gap can be controlled
experimentally by both applying an electric field perpendicular to
the graphene bilayer and tuning the inter-layer
hopping\cite{OOstinga,MacDonald}.

The next step in the study of the exciton formation would lie in
considering the long-ranged Coulomb interaction between the two
carbon layers. We conjecture that treating the long-range Coulomb
interaction might reduce the threshold $U_c$ toward a reasonable
value accessible by real graphene bilayer systems
\cite{Castro,Ohta}.

\acknowledgments{The authors are grateful to Shuichi Murakami and
Cheol Hwan Park for their comments on the manuscript.}

\appendix

\section{Diagonalization of the Hamiltonian \label{AppendixA}}

In the momentum space the bilayer Hamiltonian
\eqref{BilayerHamiltonian} reads

\begin{eqnarray*}
H = - \sum_k \psibar_k H_k \psi_k ,
\end{eqnarray*}
where $\psi^T_k = (  a_k ~ b_k ~c_k ~ d_k )$ and $H_k =$

{\small
\begin{eqnarray*}
\left(\begin{array}{cccc}
V & t\sum_\alpha e^{ik \cdot e_\alpha} & 0 & 2t_\perp \\
t\sum_\alpha e^{-ik \cdot e_\alpha} & V & 0 & 0 \\
0 & 0 & -V  & t\sum_\alpha e^{ik \cdot e_\alpha} \\
2t_\perp & 0 &  t\sum_\alpha e^{-ik \cdot e_\alpha} & -V
\end{array} \right).
\end{eqnarray*}
}
The unitary matrix diagonalizing the Hamiltonian is given by a
string of matrices,

\begin{eqnarray*}
U_k &=&  U_{0k} U_{1k} U_{2k} U_{3k},\nn
U_{0k} & = & \frac{1}{\sqrt{2}} \left(
\begin{array}{cccc}
e^{i\phi_k} & e^{i\phi_k} & 0 & 0 \\
1 & -1 & 0 & 0 \\
0 & 0 & e^{i\phi_k} & e^{i\phi_k} \\
0 & 0 & 1 & -1
\end{array}
\right), \nn
&& e^{i \phi_k } = \mathrm{ph} \left( \underset{\alpha}{\sum} e^{i
k\cdot e_\alpha}  \right).
\end{eqnarray*}
After diagonalizing with $U_{0k}$ one has $U^\dag_{0k} H_k U_{0k} =$

\begin{eqnarray*}
\left(\begin{array}{cccc}
V \!+\! \varepsilon_k  & 0 & t_\perp e^{-i\phi_k}   & -t_\perp e^{-i\phi_k} \\
0 & V \!-\! \varepsilon_k &  t_\perp e^{-i\phi_k} & - t_\perp e^{-i\phi_k} \\
t_\perp e^{i\phi_k} & t_\perp e^{i\phi_k} & -V\! +\! \varepsilon_k & 0 \\
-t_\perp e^{i\phi_k} & -t_\perp e^{i\phi_k} & 0 & -V\! -\! \varepsilon_k
\end{array}
\right).
\end{eqnarray*}
The second rotation is implemented by $U_{1k}=$

{\small
\begin{eqnarray*}
&& \left(
\begin{array}{cccc}
\cos \theta_1 & 0 & -\sin \theta_1 e^{-i\phi_k} & 0 \\
0 & \cos \theta_1 & 0 & \sin \theta_1 e^{-i\phi_k} \\
\sin \theta_1 e^{i\phi_k} & 0 & \cos \theta_1 & 0 \\
0 & -\sin \theta_1 e^{i\phi_k} & 0 & \cos \theta_1
\end{array}
\right), \nn
&&\cos 2\theta_1 = \frac{V}{\sqrt{V^2\!+\!t_\perp^2}}, ~~ \sin
2\theta_1 =\frac{t_\perp}{\sqrt{V^2\!+\!t_\perp^2}}.
\end{eqnarray*}
}
After diagonalizing with $U_{1k}$ one has

{\footnotesize
\begin{eqnarray*}
&& U^\dag_{1k} U^\dag_{0k} H_k U_{0k} U_{1k} =  \nn
&&\left(\begin{array}{cccc} \varepsilon_k \!+\! \lambda  & t_\perp^2
/\lambda & 0  &
- e^{-i\phi_k} V t_\perp  /\lambda \\
t_\perp^2/\lambda &  - \varepsilon_k \!+\! \lambda  &
e^{-i \phi_k} V t_\perp/\lambda & 0 \\
0 & e^{i\phi_k} V t_\perp /\lambda & \varepsilon_k -\lambda & t_\perp^2/
\lambda \\
- e^{i\phi_k} V t_\perp /\lambda & 0 & t_\perp^2/\lambda & -
\varepsilon_k \!-\!\lambda
\end{array}
\right),
\end{eqnarray*}
}
where $\lambda = \sqrt{V^2 \!+\! t_\perp^2}$.
The next step in the diagonalization is affected by

\begin{eqnarray*}
&& U_{2k} = \left(
\begin{array}{cccc}
\cos \theta_{2k} & -\sin \theta_{2k} & 0 & 0 \\
\sin \theta_{2k} & \cos \theta_{2k} & 0 & 0 \\
0 & 0 & \cos \theta_{2k} & -\sin \theta_{2k} \\
0 & 0 & \sin \theta_{2k} & \cos \theta_{2k}
\end{array}
\right)  ,\nn
&& \cos 2\theta_{2k} = \frac{\varepsilon_k \lambda}
{\sqrt{t_\perp^4\!+\!\varepsilon_k^2 \lambda^2}}, ~~ \sin
2\theta_{2k} = \frac{t_\perp^2} {\sqrt{t_\perp^4 \!+\!
{\varepsilon_k }^2 \lambda^2 }}. \nn
\end{eqnarray*}
After diagonalizing with $U_{2k}$ one has

{\footnotesize
\begin{eqnarray*}
&& U^\dag_{2k} U^\dag_{1k} U^\dag_{0k} H_k U_{0k}
U_{1k} U_{2k} = \nn
&&\left(\begin{array}{cccc}  \lambda + \xi_k /\lambda  & 0 & 0 &
- e^{-i\phi_k }V t_\perp  /\lambda \\
0 &  \lambda - \xi_k /\lambda  &
e^{-i \phi_k} V t_\perp/\lambda & 0 \\
0 & e^{i\phi_k} V t_\perp/\lambda &  -\lambda + \xi_k /\lambda & 0\\
- e^{i\phi_k} V t_\perp/\lambda & 0 & 0 & - \lambda - \xi_k /\lambda
\end{array}
\right),
\notag \\
\end{eqnarray*}
}
where $\xi_k = \sqrt{\varepsilon_k^2 \lambda^2 \!+\! t_\perp^4}$.
The final step in the diagonalization is given by $U_{3k}=$

{\footnotesize
\begin{eqnarray*}
\left(
\begin{array}{cccc}
\cos \theta_{3k} & 0 & 0 & \sin \theta_{3k} e^{-i \phi_k} \\
0 & \cos \theta_{4k} & -\sin\theta_{4k} e^{-i \phi_k} & 0 \\
0 & \sin \theta_{4k} e^{i\phi_k} & \cos\theta_{4k} & 0 \\
-\sin \theta_{3k} e^{i \phi_k} & 0 & 0 & \cos\theta_{3k}
\end{array}
\right), \nn
\cos 2\theta_{3k} \!=\! \frac{\lambda^2 \!+\! \xi_k}{\sqrt{V^2 t_\perp^2
\!\!+\!\! (\lambda^2 \!\!+\!\! \xi_k )^2}},  \sin 2\theta_{3k} \!=\!
\frac{V t_\perp}{\sqrt{V^2 t_\perp^2 \!\!+\!\!(\lambda^2 \!\!+\!\! \xi_k )^2}}
\nn
\cos 2\theta_{4k} \!=\! \frac{\lambda^2 \!\!-\!\! \xi_k}{\sqrt{V^2
t_\perp^2 \!\!+\!\! (\lambda^2 \!\!-\!\! \xi_k )^2}}, \sin
2\theta_{4k} \!=\!\frac{V t_\perp}{\sqrt{V^2 t_\perp^2 \!\!+\!\!
(\lambda^2 \!\!-\!\! \xi_k )^2}}.
\notag \\
\end{eqnarray*} }
After diagonalizing with $U_{3k}$ one has

\begin{eqnarray*}
&& U^\dag_{3k} U^\dag_{2k} U^\dag_{1k} U^\dag_{0k} H_k U_{0k}
U_{1k} U_{2k} U_{3k} \nn
&& =
\left(\begin{array}{cccc}
E^{++}_k & 0 & 0 & 0 \\
0 &  E^{+-}_k & 0 & 0 \\
0 & 0 &  E^{--}_k & 0\\
0 & 0 & 0 & E^{-+}_k
\end{array}
\right),
\end{eqnarray*}
where $E^{\pm \pm}_k = \pm \sqrt{\varepsilon_k^2 + \lambda^2 + t_\perp^2
\pm 2\xi_k}$.
Combining the four unitary matrices into one, $U_k = U_{0k} U_{1k}
U_{2k} U_{3k}$, the eigenoperators are obtained as

\begin{eqnarray*}
\left(
\begin{array}{c}
a_k \\
b_k \\
c_k \\
d_k
\end{array} \right) = U_k
\left(
\begin{array}{c}
\alpha_k \\
\beta_k \\
\gamma_k \\
\delta_k
\end{array}
\right).
\end{eqnarray*}

The unitary matrix $U$ leads to the following expressions needed in
the evaluation of the exciton gap, Eq. \eqref{gap-equation} :

{\small
\ba
U_{12} (k) \Ubar_{43} (k) &=& \frac{1}{4}e^{i\phi_k} (1\!-\!\sin
2\theta_{2k})(1\! +\!\cos (2\theta_1 \! +\!2\theta_{4k} )), \nn
U_{42} (k) \Ubar_{13} (k) &=& -\frac{1}{4}e^{i\phi_k} (1\!-\!\sin
2\theta_{2k})  (1\!-\!\cos (2\theta_1 \! + \!2\theta_{4k} )),   \nn
U_{42} (k) \Ubar_{43} (k) &=& \frac{1}{4} e^{i\phi_k} (1\!-\!\sin
2\theta_{2k})  \sin(2\theta_1 \! + \!2\theta_{4k} ),   \nn
U_{12} (k) \Ubar_{13} (k) &=& -\frac{1}{4}e^{i\phi_k} (1\!-\!\sin
2\theta_{2k})  \sin(2\theta_1 \! + \!2\theta_{4k} ).\nn
\label{expression-for-U}
\ea
}

\section{Second neighbour Coulomb interaction \label{AppendixB}}

The second neighbour mean-field Coulomb interaction with $U_1=0$
can be rewritten
in terms of the operator $h_i = c^\dagger_i a_i$ and $g_i = d^\dagger_i b_i$
in the following form

\begin{eqnarray*}
V_C &=& -\frac{4}{9} U_2
\sum_{i}
\Bigg\{
\notag \\
&&
\langle
h^\dagger_{i,1} -\frac{1}{2} \left(h^\dagger_{i,2} + h^\dagger_{i,3} \right)
\rangle
\left(
h_{i,1} -\frac{1}{2} \left(h_{i,2} + h_{i,3} \right)
\right)
\notag \\
&& +
\langle
h^\dagger_{i,2} -\frac{1}{2} \left(h^\dagger_{i,1} + h^\dagger_{i,3} \right)
\rangle
\left(
h_{i,2} -\frac{1}{2} \left(h_{i,1} + h_{i,3} \right)
\right)
\notag \\
&& +
\langle
h^\dagger_{i,3} -\frac{1}{2} \left(h^\dagger_{i,1} + h^\dagger_{i,2} \right)
\rangle
\left(
h_{i,3} -\frac{1}{2} \left(h_{i,1} + h_{i,2} \right)
\right)
\notag \\
&& +
\frac{3}{4}
\langle
h^\dagger_{i,1} + h^\dagger_{i,2} + h^\dagger_{i,3}
\rangle
\left(
h_{i,1} + h_{i,2} + h_{i,3}
\right)
+ h.c.
\Bigg\}
\notag \\
&& + h \rightarrow g
\end{eqnarray*}

\end{document}